\documentclass[aps,prb,twocolumn,showpacs,floatfix,superscriptaddress]{revtex4}

\usepackage{amsmath}

\usepackage{graphicx}

\begin{document}

\title{Bloch oscillations in an aperiodic one-dimensional potential}

\author{F.\ A.\ B.\ F.\ de Moura}
\author{M.\ L.\ Lyra}
\affiliation{Departamento de F\'{\i}sica, Universidade Federal de 
Alagoas, Macei\'{o} AL 57072-970, Brazil}

\author{F.\ Dom\'{\i}nguez-Adame}
\affiliation{GISC, Departamento de F\'{\i}sica de Materiales, 
Universidad Complutense, E-28040 Madrid, Spain}

\author{V.\ A.\ Malyshev}
\affiliation{Institute for Theoretical Physics and Materials
Science Center, University of Groningen, Nijenborgh 4,  9747 AG
Groningen, The Netherlands}
\thanks{On leave from ``S.I. Vavilov State Optical Institute'',
199034 Saint-Petersburg, Russia.}

\begin{abstract}

We study the dynamics of an electron subjected to a static uniform
electric field within a one-dimensional tight-binding model with a
slowly varying aperiodic potential. The unbiased model is known to
support phases of localized and extended one-electron states
separated by two mobility edges. We show that the electric
field promotes sustained Bloch oscillations of an initial Gaussian
wave packet whose amplitude reflects the band width of extended
states. The frequency of these oscillations exhibit unique
features, such as a sensitivity to the initial wave packet
position and a multimode structure for weak fields, originating
from the characteristics of the underlying aperiodic potential.

\end{abstract}


\maketitle


\section{introduction}

The nature of one-electron eigenstates has a significant influence
on the electronic transport properties of solids. In pure periodic
systems, the one-electron eigenstates are Bloch waves which are
translational invariant and all delocalized in the thermodynamic
limit. In the absence of scattering, the system behaves as a
perfect conductor whenever the Fermi energy falls into the
conduction band. Disorder, originating from lattice imperfections,
modifies the nature of the one-electron eigenstates. For a
relatively weak disorder (of a magnitude smaller than the band
width), the states at the band center may remain
extended,~\cite{Anderson58} but looses the phase coherence at
large distances. Those states, which lie near the band edges, turn
out to be exponentially localized. Well defined energies, which
separate localized and extended states, are known as mobility
edges.~\cite{Mott67} The metallic or insulating character of the
system at zero temperature depends now on whether the Fermi level
is located within the phase of delocalized or localized states,
respectively. In the former case, the system shows a finite
conductivity, while in the latter one it behaves as a perfect
insulator. A large compared to the band width disorder localizes
all the one-electron eigenstates, and the conductivity of the
system vanishes at zero temperature.

The above picture holds for three-dimensional (3D) systems 
(for an overview see Refs.~\onlinecite{Lee85} and
\onlinecite{Kramer93}). In lower
dimensions the effect of disorder is much more dramatic. In
particular, uncorrelated disorder of any magnitude causes
exponential localization of all one-particle eigenstates in one
dimension (1D)~\cite{Mott61,Abrahams79} and weak localization in
two dimensions (2D).~\cite{Abrahams79} At the end of eighties and
beginning of nineties it was realized, however, that extended
states may survive in 1D systems when the disorder distribution is
correlated.~\cite{Flores89,Dunlap90,phillips91,adame1,adame2,%
Bellani99,Moura98,Izrailev99,Kuhl00} Thus, a short-range
correlated disorder was found to cause the extended states at
special resonance energies; they form a set of null measure in the
density of states in the thermodynamic
limit.~\cite{Flores89,Dunlap90,phillips91,adame1,adame2} Because
of that mobility edges do not exist. Oppositely, long-range
correlations in the disorder distribution support a finite
fraction of the delocalized states~\cite{Moura98,Izrailev99} and
give rise to the existence of mobility edges. In
Refs.~\onlinecite{Cressoni98,Rodriguez00,Rodriguez03}
it was argued that a nonrandom long-range inter-site coupling
represents a one more driving force to delocalize one-particle 
states in 1D~\cite{Cressoni98,Rodriguez00,Rodriguez03}
and 2D~\cite{Rodriguez03} geometries. The theoretical predictions
about suppression of localization in 1D geometry due to
correlations in the disorder distribution were recently confirmed
experimentally in semiconductor superlattices with intentional
correlated disorder~\cite{Bellani99} as well as in single-mode
wave guides with inserted correlated scatterers.~\cite{Kuhl00}

Another class of 1D models, that can exhibit an Anderson-like
localization-delocalization transition, involves a nonrandom,
deterministic potential which is incommensurate with the
underlying lattice. Several models of this type have been
extensively investigated in the literature, and the localized or
extended nature of their eigenstates has been related to general
characteristics of the incommensurate
potentials.~\cite{Grempel82,Griniasty88,Thouless,sarma,yamada} An
interesting class among them is represented by aperiodic slowly
varying potentials. The latter support the mobility edges in a
very close analogy with the standard 3D Anderson
model.~\cite{sarma}

Recently, there appeared a renewed interest in the dynamics of an
electron in crystals subjected to a uniform static electric field.
Under this condition, the electronic wave function displays
so-called Bloch oscillations,~\cite{Bloch28,Wannier,Dunlap86} the
amplitude of which is proportional to the band width. The
electronic Bloch oscillation was observed for the first time in
semiconductor superlattices~\cite{Leo92} (for an overview see
Ref.~\onlinecite{Leo98}). It is to be noticed that in bulk
materials, this coherent regime of electron motion is hard to
realize because the electron dephasing time due to
scattering on defects and phonons is usually shorter than the
period of Bloch oscillations. An exception represents
semiconductor superlattices, in which the opposite situation takes
place.  The unit cells in these materials are large enough to make
the period of Bloch oscillations shorter than the electron
dephasing time, so that several periods of oscillations can be
detected.~\cite{Leo92} A similar phenomenon of sustained
oscillations of the electromagnetic field, named photon Bloch
oscillations, have also been reported in two-dimensional wave
guide arrays and optical superlattices based on porous
silicon.~\cite{Gil04}

Recently we investigated theoretically Bloch oscillations in a 1D
disordered system with diagonal long-range correlated disorder and
found that this type of correlations in disorder does not destroy
the coherence of Bloch oscillations.~\cite{prl03} The amplitude of
the latter was found to carry information about the energy
difference between the two mobility edges. This result resembles
the one that exists for an ideal Bloch band, where the amplitude of
oscillation is proportional to the bandwidth.

In this work, we further contribute to the general understanding
of the phenomenon of electronic Bloch oscillations in non-periodic
low-dimensional systems exhibiting  mobility edges. To this end,
we will focus on the biased wave packet dynamics of a single
electron moving in a lattice with an aperiodic slowly varying
potential. As potential correlations are long-ranged, sustained
oscillations may be anticipated, similarly to our previous
results.~\cite{prl03} It should be noticed, however, that the
deterministic nature of the potential and its aperiodicity
introduce new features to the Bloch oscillations. We solve
numerically the time-dependent Schr\"{o}dinger equation and
compute the time dependence of the average electron position.
Fourier analysis is implicated to reveal regularities of the
electron motion.

\section{Model system and relevant magnitudes}

We consider a tight-binding Hamiltonian on a regular 1D open
lattice of spacing $a$ with an aperiodic slowly varying potential
and a uniform static electric field ~\cite{Dunlap86,Nazareno99}
\begin{align}
{\cal H} & = \sum_{n=1}^{N}\Big(\widetilde{\varepsilon}_{n}-e{\cal
F}an\Big)
|n\rangle\langle n| \nonumber \\
 & -J\sum_{n=1}^{N-1}\Big(|n\rangle\langle n+1|
    +|n+1\rangle\langle n|\Big)\ ,
\label{hamiltonian}
\end{align}
where $|n\rangle$ is a Wannier state localized at site $n$ with
energy $\widetilde{\varepsilon}_{n}$, $\cal F$ is the external
uniform electric field and $-e$ is the charge of the particle. The
hopping amplitude is assumed to be uniform over the entire lattice
with $J>0$. In terms of the Wannier amplitudes $\psi_{n}(t)=
\langle n|\Psi(t)\rangle$, the time-dependent Schr\"{o}dinger
equation reads~\cite{Dunlap86}
\begin{equation}
i\dot{\psi}_{n}=(\varepsilon_{n}-Fn)\psi_{n}-\psi_{n+1}-\psi_{n-1}\
, \label{Schrodinger}
\end{equation}
where we introduced the dimensionless magnitudes $\varepsilon_{n}
= \widetilde{\varepsilon}_{n}/J$ and $F = e{\cal F}a/J$. Time is
expressed in units of $\hbar/J$.

\begin{figure}[ht]
\centerline{\includegraphics[width=80mm,clip]{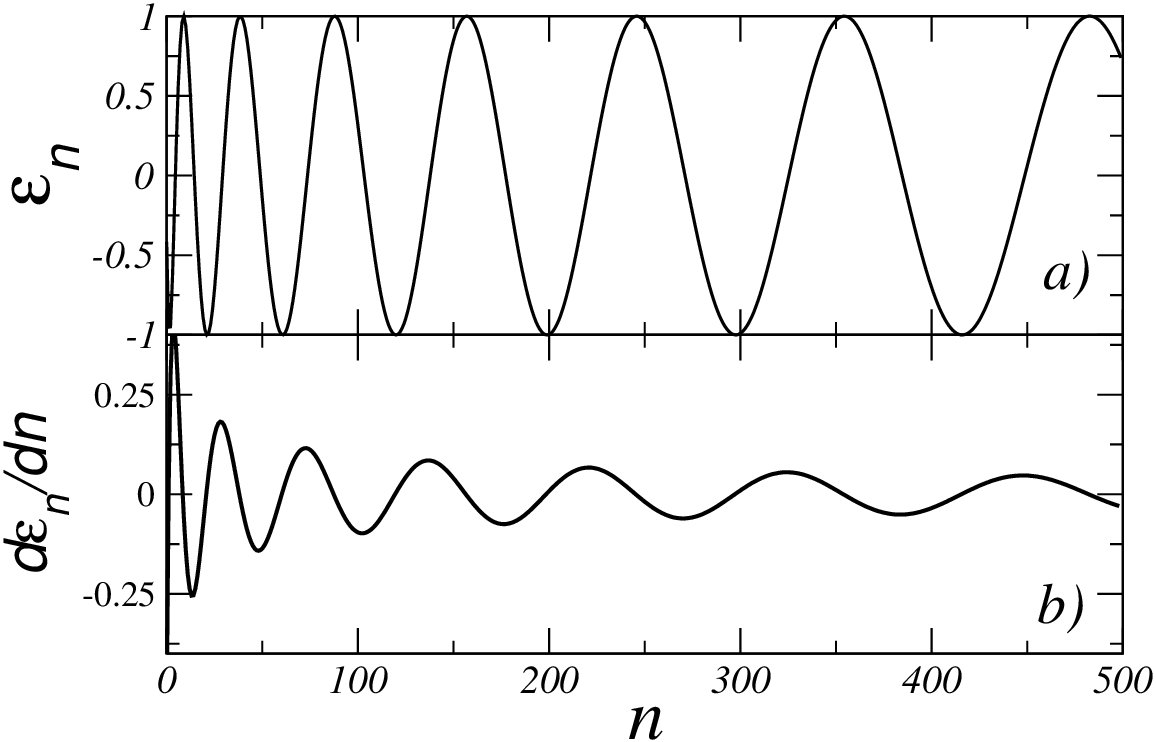}}
\caption{ (a) Aperiodic slowly varying potential calculated for $V
= 1$, $\pi\alpha = 1$, and $\nu = 0.5$; \,  b) The potential
derivative showing the decrease of its local maxima on increasing
the site index $n$.} \label{fig1}
\end{figure}

Here, we will consider an on-site potential $\varepsilon_{n}$
given by~\cite{sarma}
\begin{equation}
\varepsilon_n = V\cos\Big(\pi\alpha n^{\nu}\Big)\ , \qquad V>0 \ ,
\end{equation}
with $V$, $\alpha$ and $\nu$ being variable parameters (an example
of $\varepsilon_{n}$ is depicted in Fig.~\ref{fig1}). For $\nu =1$
this is just Harper's model, for which a rational $\alpha$
describes a crystalline solid, whereas an irrational $\alpha$
results in an incommensurate potential. Dynamical localization in the Harper's model under the 
action of electric fields was studied in Ref~\onlinecite{naza95}. 
In the following, we
restrict our study to an interesting range of parameters $0<\nu<1$
and $0<V<2$. It was demonstrated for this case that there exists a
phase of extended states near the band center, which is separated
by the two mobility edges $\pm E_c = \pm (2 - V)$ from two phases
of localized states closer to the band edges.~\cite{sarma}

The pertinent quantities we will use to characterize the dynamics
of the electron wave packet are its mean position (centroid)
\begin{subequations}
\begin{equation}
x(t)= \sum_{n=1}^{N}(n-n_0)\,|\psi_{n}(t)|^2\ , \label{tools1}
\end{equation}
and the \emph{spread} of the wave function (square root of the
mean squared displacement)
\begin{equation}
\sigma(t)= \left( \sum_{n=1}^{N}\left[n-\langle n(t)\rangle
\right]^2 \,|\psi_{n}(t)|^2\right)^{1/2} , \label{tools2}
\end{equation}
\end{subequations}
where $\langle n(t)\rangle=\sum_{i=1}^Nn|\psi_{n}(t)|^2$. As the
initial packet is assumed spatially narrow, one has contributions
to the wave packet dynamics, coming from a wide spectrum of
eigenstates of the Hamiltonian~(\ref{hamiltonian}).

In ideal lattices, a uniform field causes the electron wave packet
to oscillate in space and time. The amplitude and the period of
these oscillations are estimated semiclassically as $L_c = W/2F$
and $\tau_B=2\pi/F$, respectively (see, e.g.,
Ref.~\onlinecite{Ashcroft}), where $W$ here is the width of the
Bloch band in units of the hopping constant $J$. Subsequently,
the frequency of the harmonic motion is $\omega = F$, for the
chosen units. The above picture was shown to remain valid when
long-range correlated disorder is introduced with $W$ being
replaced by the band width $W_c$ of delocalized
states.~\cite{prl03} In what follows, we provide numerical
evidences of that the above semiclassical picture still holds for
the present aperiodic system, although this requires a
renormalization of the applied electric field due to the presence
of the on-site aperiodic potential.

\section{Numerical results}

We numerically solved the time-dependent Schr\"{o}dinger
equation~(\ref{Schrodinger}) by means of an implicit integration
algorithm.~\cite{Press86} The initial wave packet was chosen to be
a narrow Gaussian of width $\sigma=1$, centered at an arbitrary
lattice site $n_0$:
\begin{equation}
\psi_{n}(0)=
A(\sigma)\exp{\left[-\,\frac{(n-n_0)^2}{4\sigma^2}\right]}\ ,
\label{Gaussian}
\end{equation}
where $A(\sigma)$ is a normalization constant. To improve
stability of the numerical algorithm, we set the electric
potential to vanish at the initial site, replacing $Fn$ by
$F(n-n_0)$ in Eq.~(\ref{Schrodinger}). This produces only a shift
in the origin of energy with no  physical effect.

\subsection{Unbiased dynamics}

\begin{figure}[ht]
\centerline{\includegraphics[width=70mm,clip]{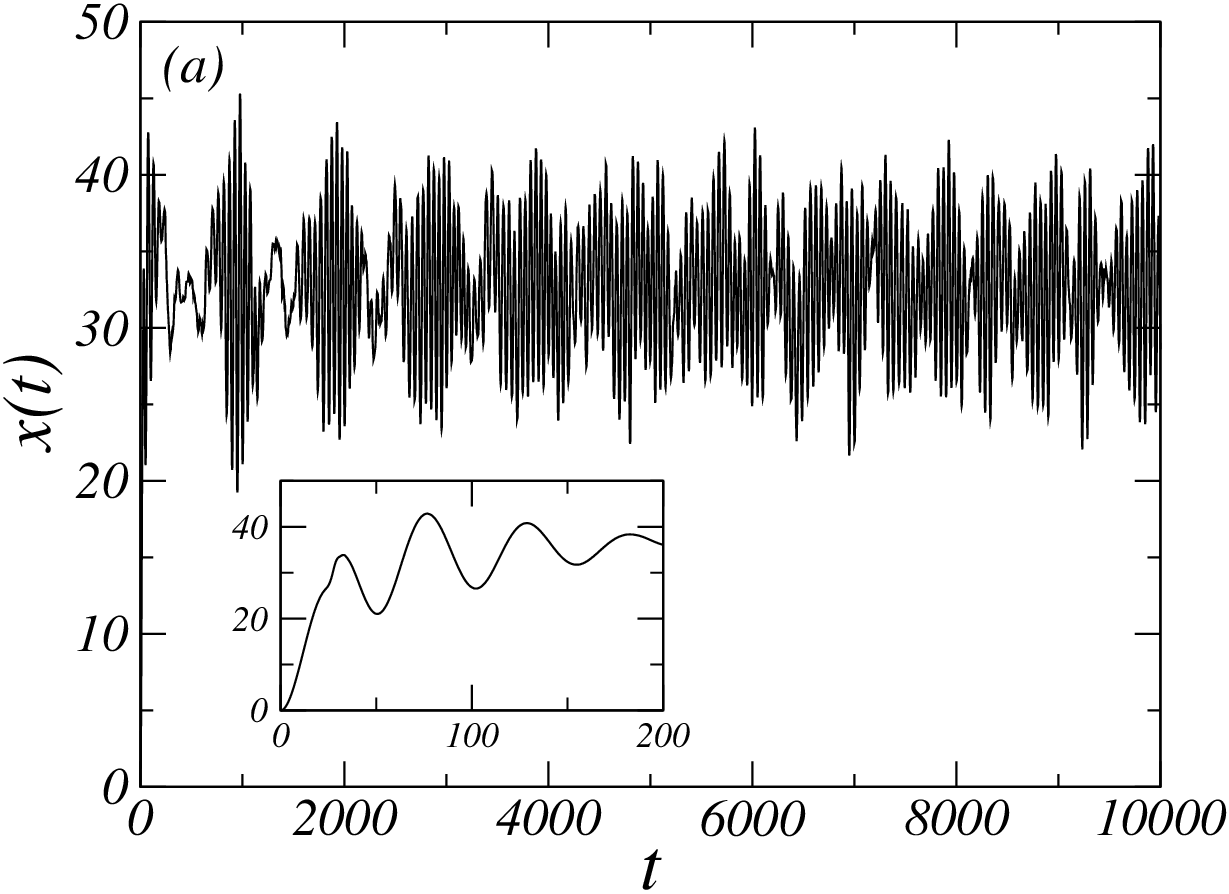}}
\centerline{\includegraphics[width=70mm,clip]{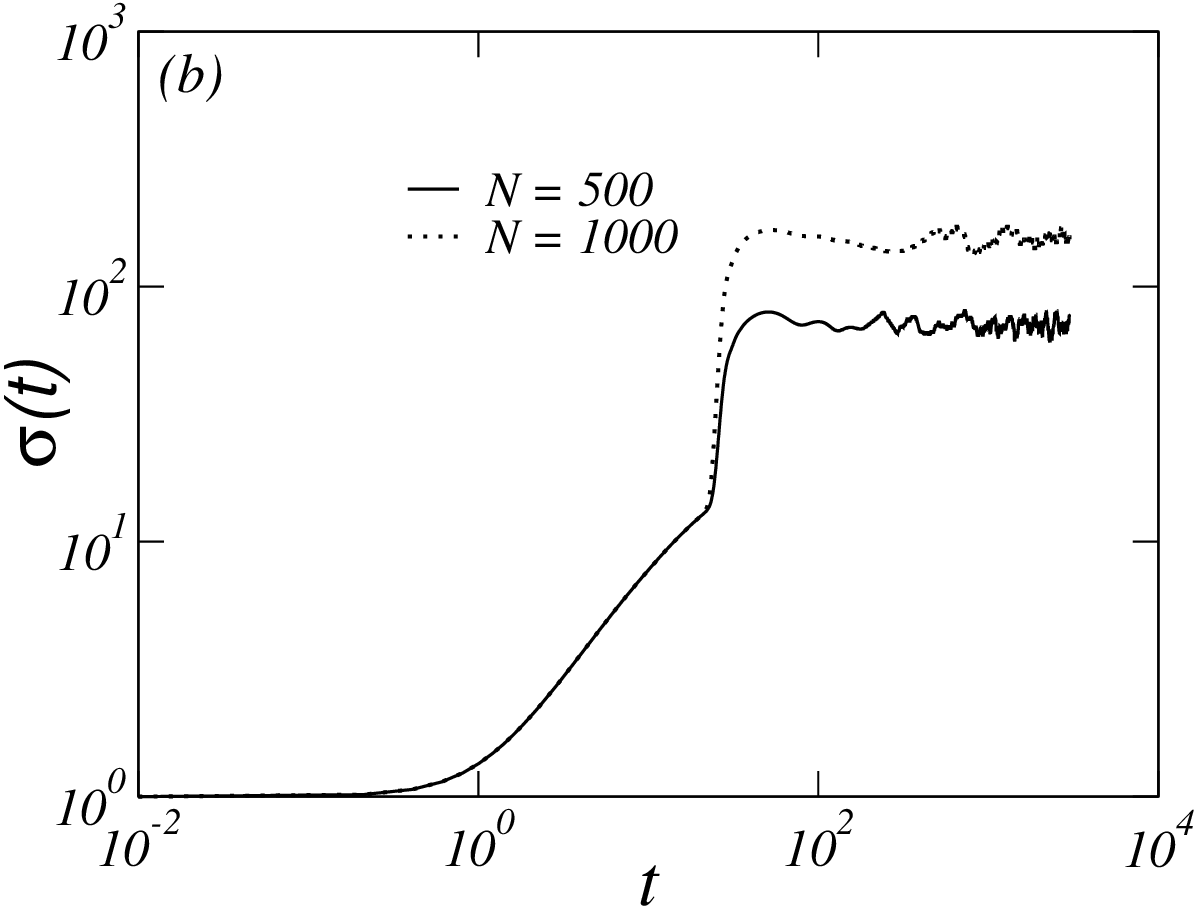}}
\caption{(a) Time-domain dynamics of the centroid of an unbiased
wave packet ($\sigma=1$, $n_0=48$  at $t = 0$). The inset shows
the coherent oscillations at short times. (b) Time-domain dynamics
of the wave function spread for two different system sizes.}
\label{fig2}
\end{figure}

We start our analysis of the electron motion by studying the wave
packet dynamics in the absence of the external field. The
parameters of the aperiodic potential are set hereafter to $V=1$,
$\pi\alpha=1$, and $\nu=0.5$. In Fig.~\ref{fig2}(a), we depicted
the time evolution of the centroid of a wave packet, which
initially was located at site $n_0=48$. It is worth to notice that
the potential slope at this point is negative (see Fig.~\ref{fig1}).
The inset in Fig.~\ref{fig2}(a) demonstrates that at the earlier
stage of motion, the centroid oscillates harmonically, but loses
the phase memory rapidly. The frequency of these few oscillations
is about $0.13$ in the dimensionless units. We show below that this
number is directly related to the strength of the local field,
produced by the aperiodic potential in the vicinity of the initial
position of the centroid, $n_0 = 48$. Thus, the oscillations found
can be referred as to {\it zero-field} Bloch-like oscillations. To
prove this statement, one should bear in mind that at short times
the spread is small on the scale of the aperiodic potential [see
Fig.~\ref{fig2}(b)], so that the latter can be represented by two
first terms of its Tailor expansion around the initial position of
the centroid. Then, the potential derivative $d\varepsilon_n/dn$
determines the local field strength which, in the dimensionless
units, should be interpreted as the frequency of zero-field
Bloch-like oscillations, similarly to the biased Bloch oscillations
(see the preceding Section). Applying this reasonings to the site
$n_0 = 48$, we find that the potential derivative is about $-0.14$
at this site [see Fig.~\ref{fig1}(b)], whose absolute value is
indeed fairly close to the frequency of oscillations $0.13$
extracted from the centroid dynamics. This is an unambiguous
confirmation of our qualitative picture. It is to be noticed,
however, that these arguments fail in the vicinities of potential
maxima and minima, where the potential slope is vanishingly small.

\begin{figure}[ht]
\centerline{\includegraphics[width=80mm,clip]{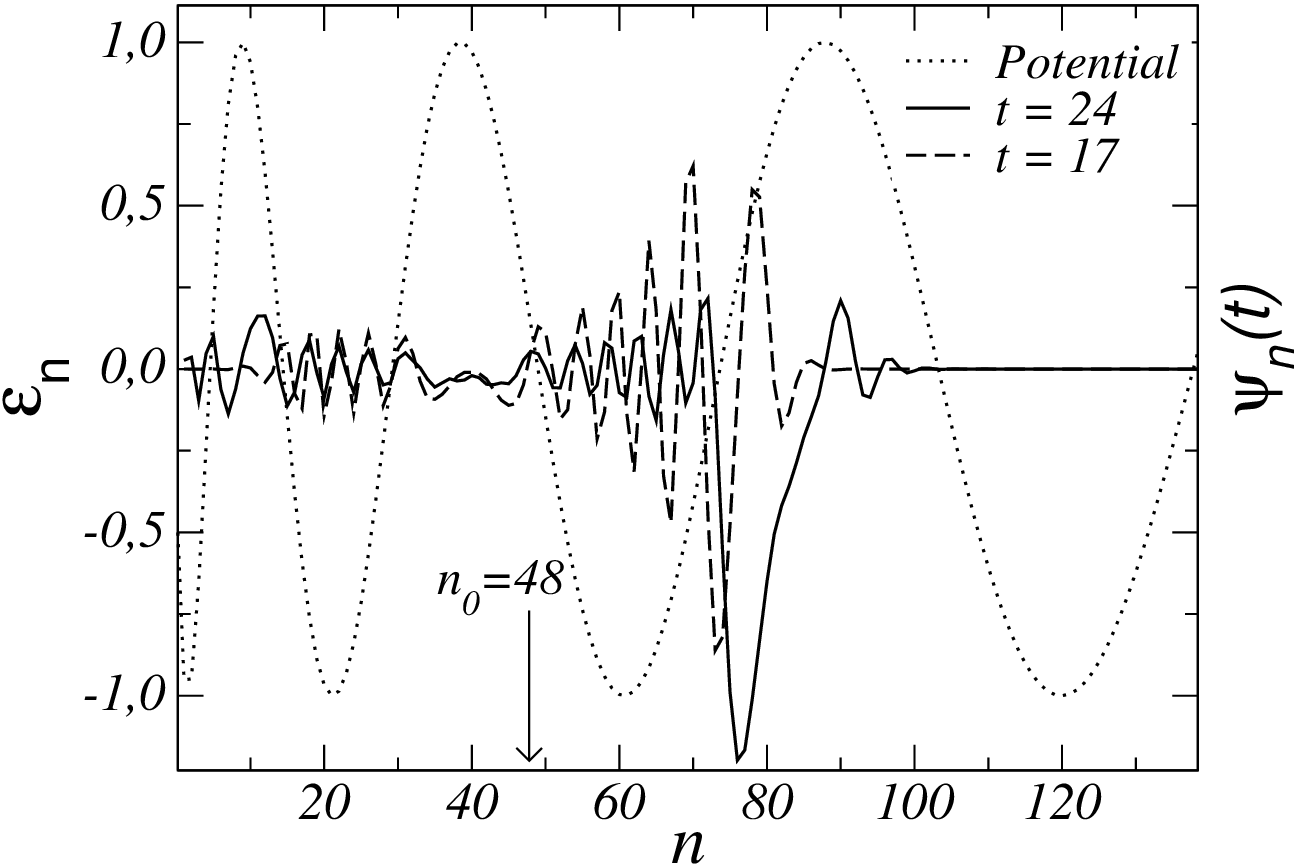}}
\caption{Wave packet profile $\Psi_n(t)$ (in arbitrary units)
calculated for two moments of time, displaying the tunneling
process from the initial potential well. All parameters are the
same as in Fig.~\protect{\ref{fig2}}.} \label{fig3}
\end{figure}

As was already mentioned, the zero-field oscillations are not
sustained for a long time. When the wave packet spread becomes
comparable with the local scale of the aperiodic potential, the
nature of the latter, with its wells and barriers, comes into play.
This results in loosing  the phase memory of the wave function. In
Fig.~\ref{fig2}(b) we depicted the time-domain behavior of the wave
packet spread. It displays a ballistic regime within some time
interval presented by a linear in log-log scale dependence,
followed further by a saturation at some value which, from one
hand, depends on the system size, but from the other hand, is
considerably smaller than that size. One can also see that the wave
packet spread reveals a steep increase at some instant of time
which is almost size independent. We relate this features to the
tunneling of the wave packet from the potential well, where it was
initially located, to the adjacent one. Figure~\ref{fig3}
illustrates this tunneling process.

\subsection{Bias effects}

\begin{figure}[ht]
\centerline{\includegraphics[width=65mm,clip]{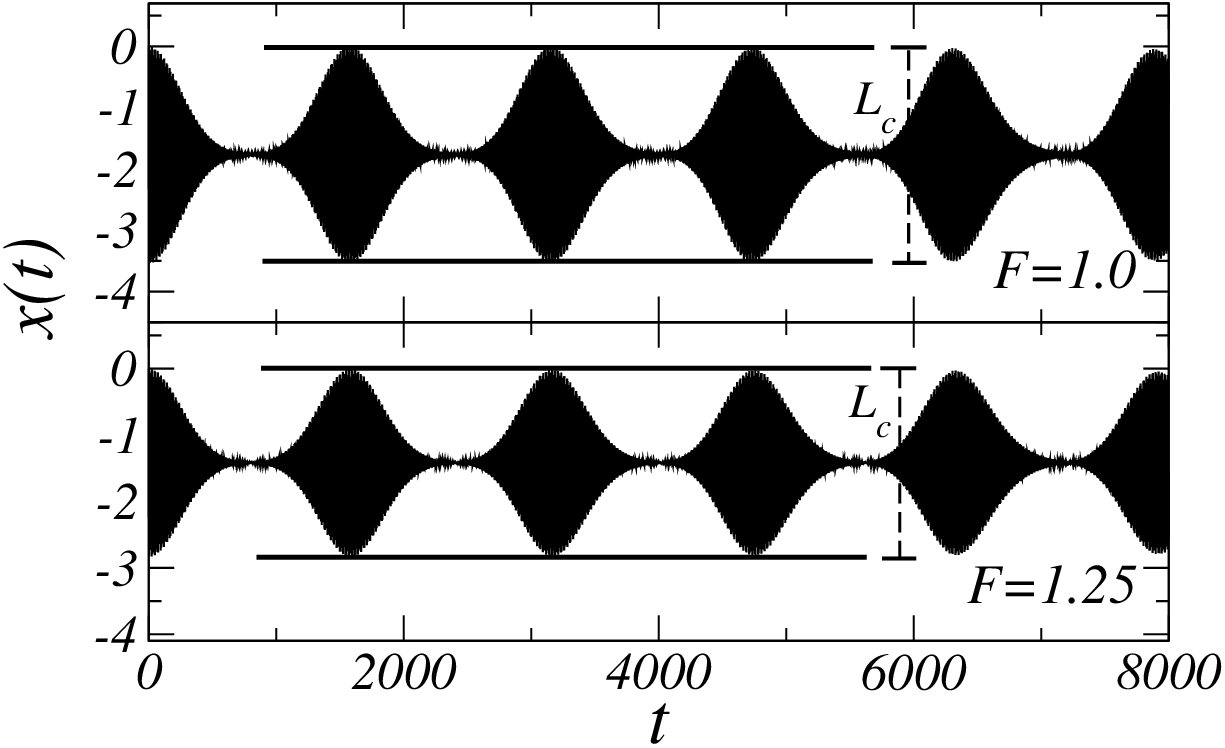}}
\centerline{\includegraphics[width=65mm,clip]{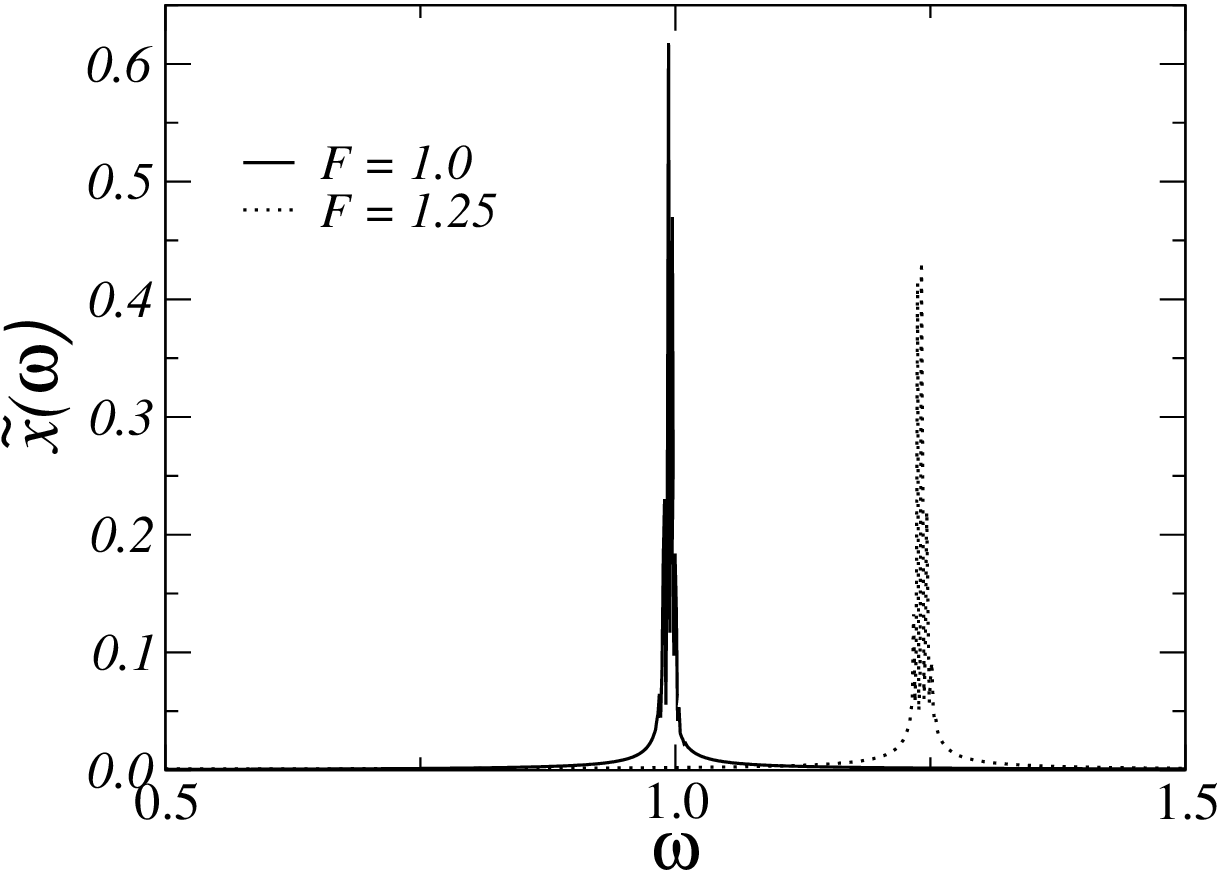}}
\caption{(a) Time-domain dynamics of the centroid of a biased wave
packet ($n_0=N/2$ and $\sigma=1$ at $t = 0$) for two values of the
applied electric field $F=1$ and $1.25$. (b)~Fourier transform of
the centroid.} \label{fig4}
\end{figure}

\begin{figure}[ht]
\centerline{\includegraphics[width=70mm,clip]{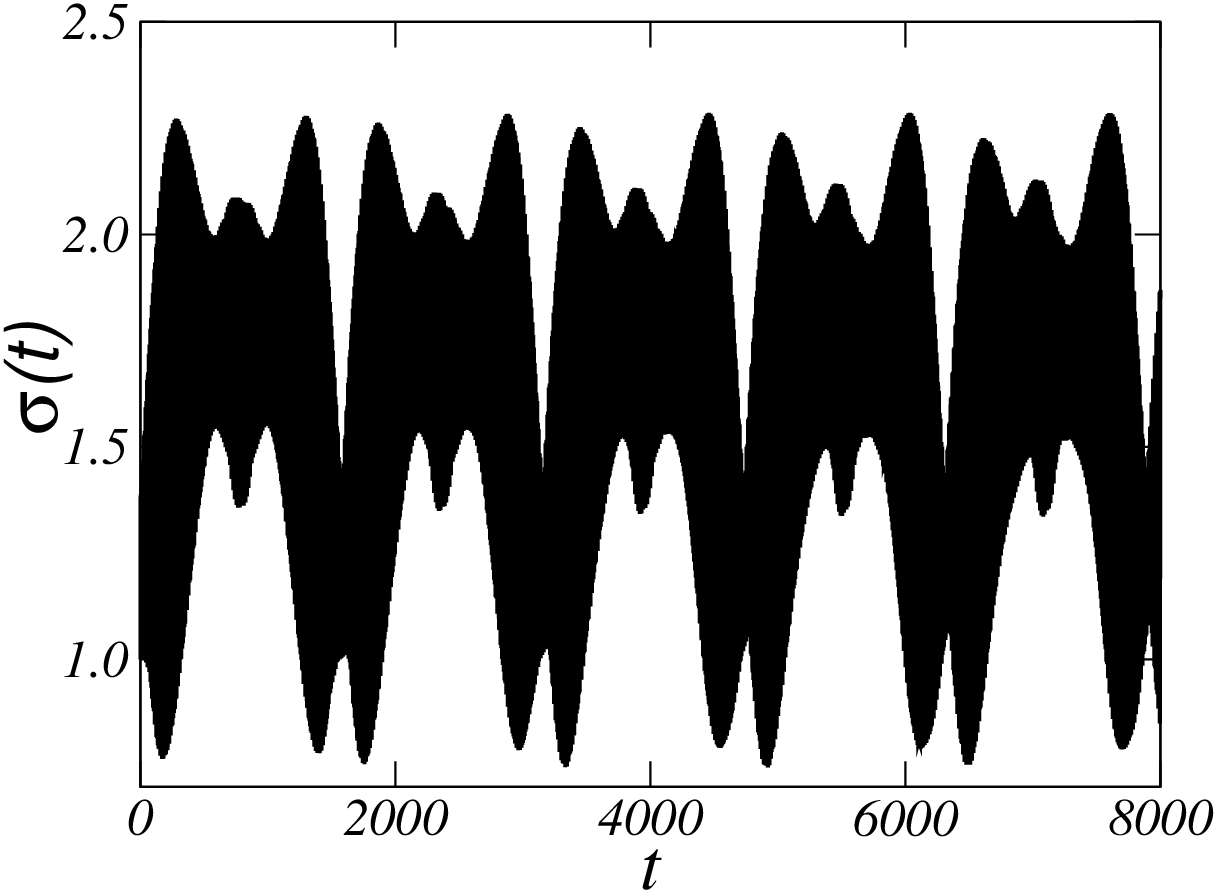}}
\caption{Time-domain dynamics of the wave function spread of a
biased wave packet ($n_0=N/2$ and $\sigma=1$ at $t = 0$). Notice
its localized and breathing nature.} \label{fig5}
\end{figure}

Once a uniform static electric field is turned on, the wave packet
dynamics presents quite distinct features as compared to the
unbiased behavior. In Fig.~\ref{fig4}(a) we plotted the centroids
of wave packets calculated for distinct field strengths for the
initial position at the chain center. It should be noticed that
now Bloch oscillations remain sustained, i.e., no dephasing is
taking place. Second, the oscillation amplitude is proportional to
$1/F$ as predicted semiclassically. To provide further
confirmation of the semiclassical picture, we calculated
numerically the Fourier transform of the centroid,
$\widetilde{x}(\omega)$, as shown in Fig.~\ref{fig4}(b). Again,
the estimated predominant frequency of the Bloch oscillations
$\omega = F$ is corroborated. The localized character of the wave
packet is revealed in Fig.~\ref{fig5} where the spread of the wave
function is shown; it saturates at a finite value independent of
the chain size, displaying however periodic oscillations. The
latter characterize the breathing nature of the oscillating wave
packet.

As already mentioned, the oscillations are not damped but remain
amplitude-modulated. Solid lines in Fig.~\ref{fig4}(a) bound  the
spatial region within which the wave packet oscillates for a long
time. The extent of this region $L_c$ is found to be $L_c \sim
W_c/F$, where $W_c$ is independent of the applied field $F$. From
the data in Fig.~\ref{fig3} we obtain $W_c \sim 2E_c$. This value
agrees remarkably well with the width of the band of extended
states reported in Ref.~\onlinecite{sarma}. Thus, we arrive at one
of the main conclusions of this work, namely there exist clear
signatures of Bloch oscillations of a biased Gaussian wave packet
between the two mobility edges.

\begin{figure}[ht]
\centerline{\includegraphics[width=70mm,clip]{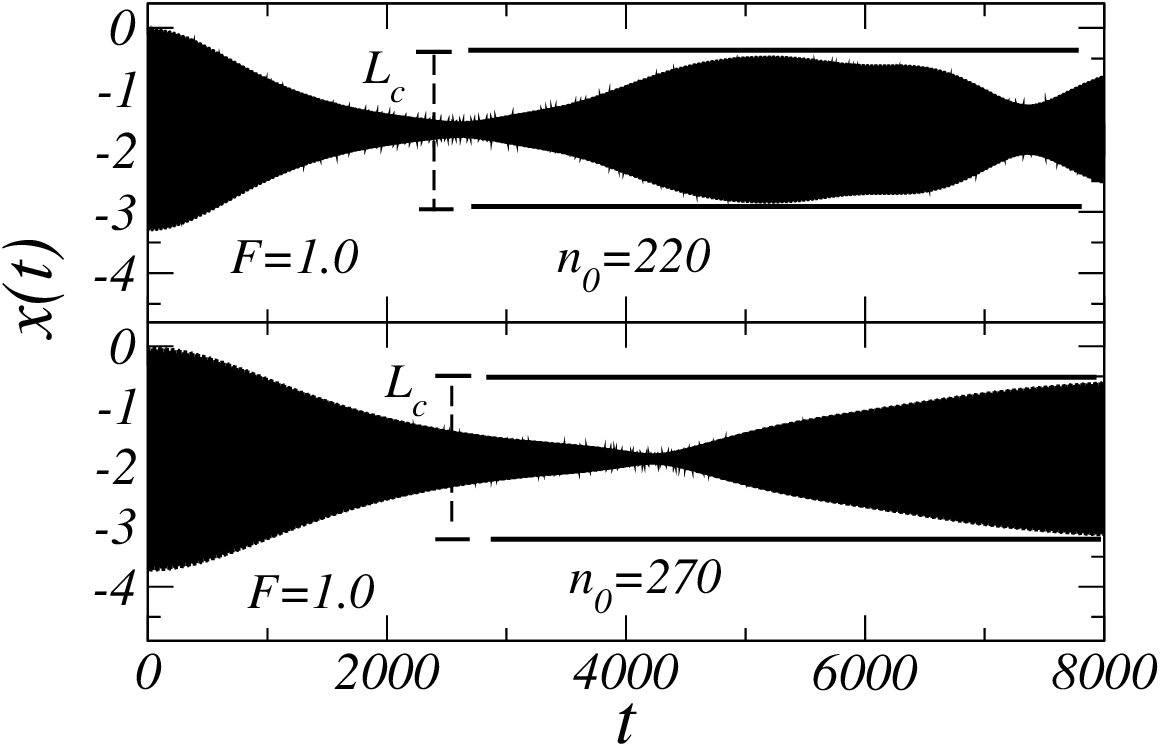}}
\centerline{\includegraphics[width=70mm,clip]{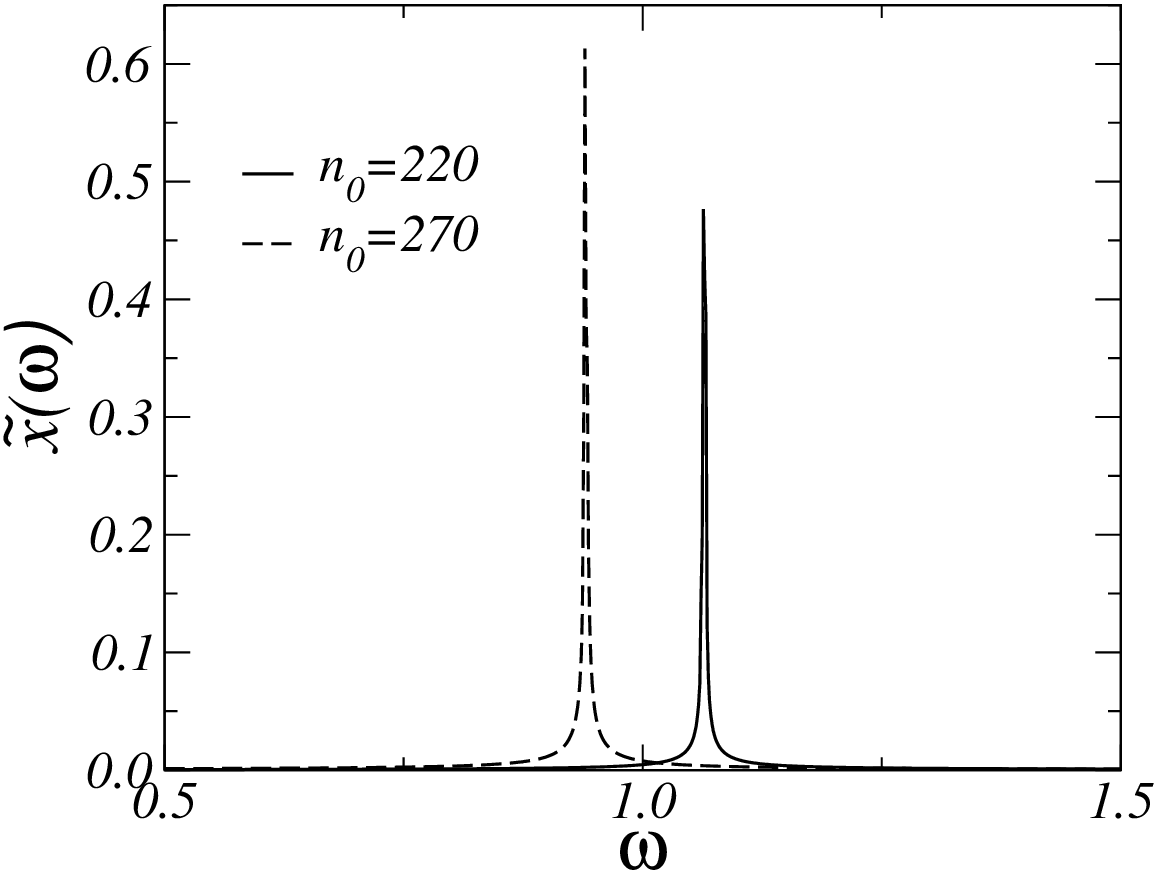}}
\caption{(a) Time-domain dynamics of the centroid of a biased wave
packet ($\sigma=1$  at $t = 0$) for $F=1$ and two different
initial positions. (b)~Numerical Fourier transform of the
centroid.} \label{fig6}
\end{figure}

\begin{figure}[ht]
\centerline{\includegraphics[width=70mm,clip]{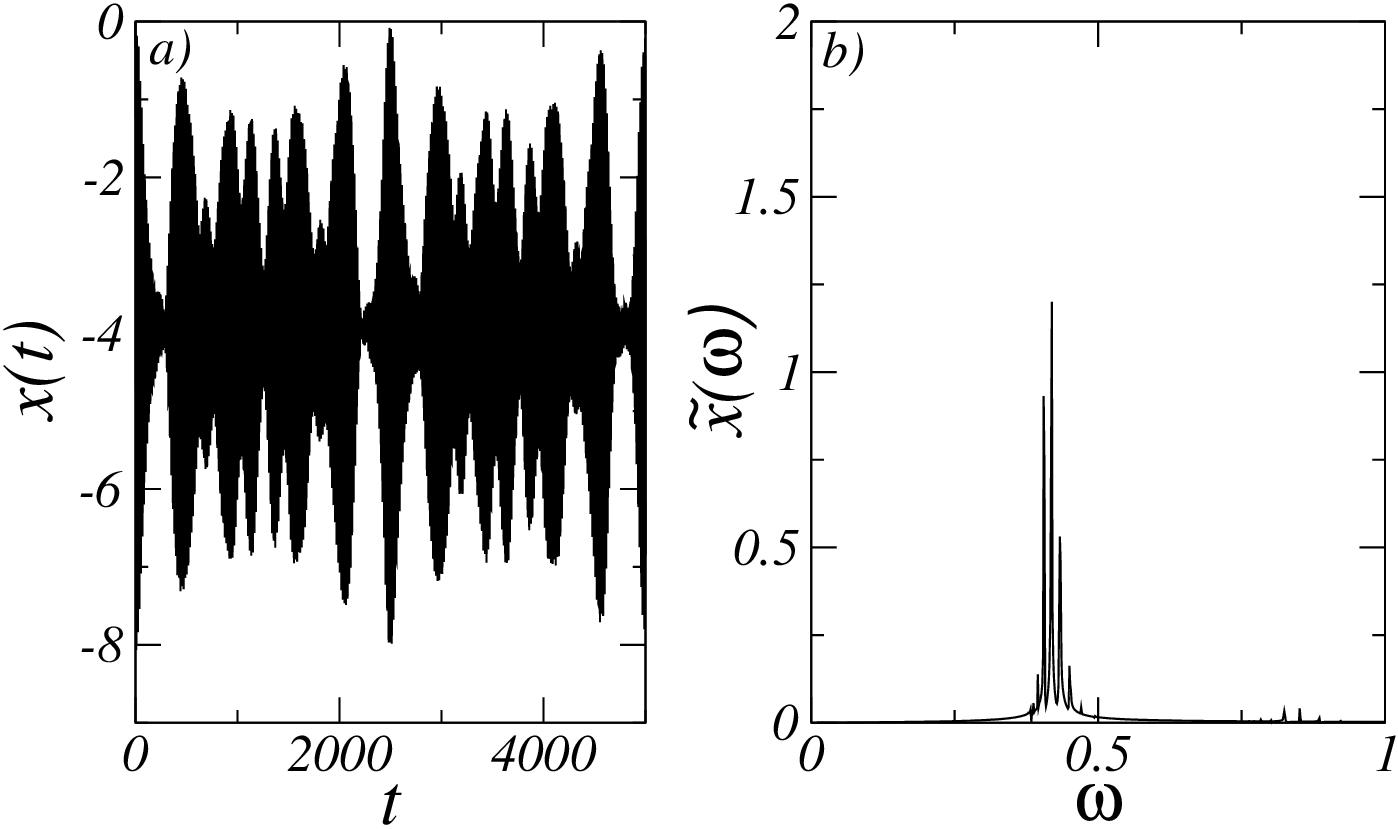}}
\caption{(a) Time-domain dynamics of the centroid of a biased wave
packet ($n_0=48$ and $\sigma=1$ at $t = 0$) and $F=0.5$.
(b)~Numerical Fourier transform of the centroid. The multi-mode
structure reveals the spread of the wave function on multiple
potential wells.} \label{fig7}
\end{figure}

Some of the above characteristics of the Bloch oscillations can
depend on the initial location of wave packets. In particular, at
the chain center the underlying aperiodic potential has an almost
null derivative. Under this condition, the bias is solely due to
the external field. In general, there is a contribution to the
external field coming from the aperiodic potential; the total
field at position $n_0$ is given by
$F_\text{eff}=d\varepsilon_n/dn + F$. The local contribution to
the applied electric field shall be relevant whenever the
potential gradient (in appropriate units) is of the order of the
applied field. This is illustrated in Fig.~\ref{fig6}(a) with wave
packets starting at locations with positive and negative potential
gradients. One can see that the oscillation amplitude becomes
larger near regions with positive gradients and smaller for
negative ones as compared with the results obtained in
Fig.~\ref{fig4} for the same applied electric field. Further, the
characteristic frequency of these oscillations also exhibits
shifts which are of the order of the local potential gradient [see
Fig.~\ref{fig1}(b)], as depicted in Fig.~\ref{fig6}(b).

As a final remark, the Bloch oscillations shall develop a more
complex structure for wave packets initially located near the
beginning of the chain, especially at small electric fields for
which the amplitude of the oscillations is large enough. Then the
wave packet can probe regions with distinct scales of the
potential gradient. In Fig.~\ref{fig7} we show a particular case
of the dynamics of a wave packet starting at $n_0=48$ under the
applied field $F=0.5$. Notice the rich pattern of the centroid
oscillations. Its Fourier transform exhibits multi-mode structure
with a series of narrow peaks which are shifted from $\omega=F$ by
quantities that are of the order of the maximum potential
gradients at the region covered by the wave packet. The shifts to
smaller frequencies can be much easily detected once the initial
packet is concentrated at a region of negative potential gradient.
However, the corresponding shifts to higher frequency can still be
detected too as a series of small peaks.

\section{Summary and concluding remarks}

We studied theoretically the single-electron wave packet dynamics
within a tight-binding model with an aperiodic slowly varying site
potential under the influence of a static uniform electric field.
The on-site potential parameters were chosen such that, in the
absence of the electric field, the model supports a phase of
delocalized states at the center of the band, similarly to high
dimensional disordered systems. The electric field promotes a bias
which localizes the electron states. The resulting wave packet
dynamics reveals Bloch oscillations. However, contrary to what
occurs in disordered systems, where scattering on site potential
fluctuations gradually degrades the oscillations, these remain
sustained with no signature of depletion.

The slowly varying aperiodic nature of the potential results in an
additional local bias. Thus, the total bias has contributions of
both the aperiodic potential and the one of the applied field. By
defining an effective local field, we could show that the
amplitude of the oscillations agrees well with the semiclassical
prediction and can be used to estimate the width of the band of
extended states. The typical frequency of these oscillations can
also be understood on the semiclassical grounds. In particular, it
is shifted depending on the local potential gradient. In the week
field limit, these oscillations exhibit a multi-mode structure as
the wave packet probes a larger region of the aperiodic potential.

Our findings indicate that Bloch oscillations can indeed be
observed in superlattices with slowly varying periodicity. The
richness of the predicted dynamical behavior can lead to new
electro-optical devices, aiming to explore the coherent motion of
confined electrons. We hope that the present work will stimulate
experimental activities along this direction.

\acknowledgments

We are very thankful to H. Yamada for bringing to our attention
the problem of aperiodic potentials with mobility edges. Work at
Brazil was supported by CNPq and CAPES (Brazilian research
agencies) and FAPEAL (Alagoas State agency). Work at Madrid was
supported by DGI-MCyT (MAT2003-01533).

\end{document}